\documentstyle[epsf,epsfig,preprint,amsmath,aps,amssymb]{revtex}

\begin{document}

\draft\preprint{SNUTP--02/036}
\title{\Large\bf Softening of SUSY 
flavor problem from asymptotically free bulk interactions} 
\author{Kang-Sin Choi\footnote{ugha@th.physik.uni-bonn.de},
Ki-Young Choi\footnote{ckysky@th.physik.uni-bonn.de} 
and Jihn E. Kim\footnote{jekim@phyp.snu.ac.kr,
jekim@th.physik.uni-bonn.de}
} 
\address{Physikalisches Institut, Universit\"at Bonn,\\
\vskip -0.3cm Nussallee 12, D53115, Bonn, Germany, and\\
School of Physics and Center for Theoretical Physics,\\
\vskip -0.3cm Seoul National University, Seoul 151-747, Korea
}
\date{\today}

\maketitle

\begin{abstract}
Recently, it was pointed out that soft masses of
the supersymmetric gauge theories with extra dimensions
tends to a flavor conserving point, which is a desirable
scenario in gravity mediation models. We point 
out that in 6D we must consider the anomaly free condition
in addition to the condition on the asymptotic freedom. 
From this, we find $E_6$,  $E_7$ and $E_8$ are natural 
candidates in 6D. There is no $SU(N)$ model, but there
exist two $SO(10)$ models and $SO(2n)$ models(one each
for each $n\ge 6$) satisfying these conditions.
In 5 dimensions,
there is no such condition on anomaly freedom, 
but the softening may not be enough.  
\\
\noindent [Key words: SUSY flavor problem, extra dimensions, 
 6D gauge anomaly]  
\end{abstract}


\def\p{\partial}
\def\L{\Lambda}
\def\MG{M_{GUT}}
\def\Mg{$M_{GUT}$}
\def\Ms{$M_{s}$}
\def\Mp{$M_{P}$}

\newpage

\section{Introduction}

The flavor problem in the gravity mediation scenario 
has to be resolved if it is going to descibe 
the soft terms of the minimal supersymmetric
standard model(MSSM) successfully. In the 4 dimensional
(4D) supergravity models, this has been known to be an
extremely difficult problem\cite{dn}. With 
the advent of new tries on extra dimensions\cite{kawamura},
this flavor problem can be reconsidered toward a possible
understanding of the SUSY flavor problem.
 
The recent $\lq$extra dimensional scenarios' are based on
the hope that these extra dimensional field theories
are obtainable from compactifications of 10D superstring
models or 11D M-theory\cite{6d}. In the early string models, 
it was argued\cite{kaplunovsky} that the GUT scale \Mg,
the string scale \Ms, and the reduced Planck scale \Mp\  
are considered to be of the same order,
under the assumption that a 4D SUSY field theory is
obtained from a 10D SUSY field theory which  in turn is 
considered to be a valid effective
theory below the string scale $M_s$. However, the
string scale of order $6\times 10^{17}$ GeV\cite{kaplunovsky} 
is known to be somewhat larger than the unification scale
$\MG\sim 2\times 10^{16}$ GeV determined from the renormalization 
group running of the observed low energy couplings.

Contrary to this early prediction on O(1) number for
the mass ratios, phenomenologically 
we need to introduce a small parameter,
\begin{equation}\label{small}
\frac{\MG}{M_P}\simeq 10^{-2}
\end{equation} 
where $M_P\simeq 2.44\times 10^{18}$ GeV is
the reduced Planck mass.  
Initially, this small number has been considered to be
a problem in perturbative string models. Therefore, 
Horava and Witten proposed
a relatively large $11^{\rm th}$ dimension with two 
9-branes with an $E_8$ group at each brane to
interpret this small number\cite{hw}.

Recently, Friedmann and Witten\cite{fw} estimated \Mg\ from
the top-down approach with the 11D supergravity compactified
with a $G_2$ holonomy. In this top-down calculation, they seem
to obtain a small number if $\alpha_{GUT}\simeq 1/25$,
\begin{equation}\label{g2}
\left(\frac{\MG}{M_P}\right)\simeq \alpha_{GUT}^{3/2} 
\times\frac{L(Q)^{1/3}}{\sqrt{4\pi a}} 
\end{equation}
where $a$ is an appropriate ratio of the 7D compact internal 
space and 7/3 power of the 3D internal space of 7D supergravity,
and $L(Q)$ is the O(1-10) number of the lens space.
Numerically, then Eq. (\ref{g2}) turns up a number 
of O($10^{-2}$). 
Even though one can argue that the 
Friedmann-Witten calculation (\ref{g2}) is for
a specfic model, it may have some truth in it if
the volume of the extra dimension is relatively large.
One notable difference of this calculation from
that of Ref.\cite{kaplunovsky} is that a 10D SUSY
field theory is not considered as an intermediate effective
theory. The recent tries 
of the extra dimensional field theories
also do not assume a 10D SUSY field theory.

In Ref.\cite{kaplunovsky}, it was pointed out in
addition that for a 6D SUSY field theory
between the string scale and the GUT scale($\simeq$ 
compactification scale), one has
$M_s/\MG<O(1/\sqrt{\alpha_{gut}})$. If $\alpha_{gut}\sim
1/25$, then the scale \Mg\ can be at most $0.1M_s$,
which was the reason that Ref.\cite{kaplunovsky}
assumed that even in 6D a small number (\ref{small})
is unreasonable. However,
with a power-law asymptotic freedom above the scale \Mg,
$\alpha_{gut}$ can be much smaller
than $\frac{1}{25}$ and a large discrepancy
between \Mg\ and $M_s$ can be generated. This
power law running was not used in Ref.\cite{kaplunovsky}.    

With a large internal space volume, many Kaluza-Klein(KK) modes
in the bulk can contribute significantly in
the running of the gauge
couplings, leading to a power law instead of a 
logarithmic running\cite{dienes}. If an effective 
4D $\beta$ function contributed by the bulk fields 
is negative, the corresponding gauge coupling constant 
decreases very rapidly at shorter distance scales. This 
can be translated to a ratio between the compactification
volume and an appropriate Planck scale.
Thus, if the volume of the internal space is large
compared to a Planckian volume, there is
a chance to understand the small number (\ref{small}). 
If the SUSY flavor problem is related to this small
number, there is a hope to understand it with extra 
dimensions.

Indeed, Kubo and Terao\cite{kubo} 
investigated the possibility of solving
the SUSY flavor problem using the
small number (\ref{small}) without introducing bulk
matter. In this paper, we confirm their conclusion 
even with bulk matter, and obtain several candidate GUT
groups in 6D from the conditions on the asymptotic freedom
and anomaly freedom.

\section{The Kubo-Terao mechanism}

Let us briefly discuss the Kubo and
Terao(KT) idea\cite{kubo} in higher dimensional SUSY field
theory models, notably in 6D models. Here, gauge multiplets 
are put in the bulk and matter multiplets are put only at the 
branes. To realize this kind of setting from 
string theory, the compactification creates matter
only at the branes.\footnote{Indeed, there exists  
an example close to this requirement 
in an orbifold compactification\cite{kimkim}.}
However, we argue that it is not an absolute requirement
to put matter only at the branes. An asymmetric assignment
of matter in the bulk and branes can be more flexible
in understanding top-bottom mass hierarchy\cite{khdkl},
and still a kind of KT mechanism can work, since the
essence of the KT mechanism is the asymptotic freedom
of the gauge couplings in the bulk and the existence of KK towers
from the bulk fields. Note that 
the threshold effect of Ref.\cite{fw}
relies only on the topology of the internal space, not
needing a knowledge on the KK spectrum, which
made it easy to write the answer in the simple form
given in (\ref{g2}).\footnote{ 
At present, it is not known how to apply the KT mechanism
in the $G_2$ holonomy case since a detail knowledge on the
KK spectrum in the bulk is not needed in this case.} 

The orbifold compactifications toward 4D and 6D models are 
extensively tabulated in the literature\cite{table}. Two
explicit 6D models ($SO(16)$ and $E_7$) 
are obtained by a $Z_2$ orbifold compactification\cite{6d}.
Here we study SUSY field theories in 5 or 6 dimensions,
but with a keen eye on possible compactifications from
10D string theory or 11D M-theory.

In this paper, we assume that below the string 
scale $M_s$ particle 
interactions are effectively described by a
$(4+\delta)$--dimensional field theory with $\delta$ a
small number. Specifically, we will choose $\delta=2$.
$M_s$ may or may not
coincide with the Planck mass $M_P=2.44\times 10^{18}$ GeV,
but it is known to be close to $M_P$\cite{kaplunovsky} and
we take this viewpoint.
We also assume for simplicity that the grand unification
scale \Mg\ is the KK scale $\equiv \frac{1}{R}$ where $R$
is the compactification radius in the sense that the compact
volume is $X_\delta R^\delta$ with\footnote{Indeed, 
if a GUT group is broken geometrically as in
Ref.~\cite{kawamura}, \Mg\ is the scale of the first 
KK mass.}
\begin{equation} 
X_\delta=
\frac{\pi^{\delta/2}}{\Gamma(1+\delta/2)}
\rightarrow (2,\pi,\frac{4\pi}{3})\ 
{\rm for\ }\delta=1,2,3. 
\end{equation}
Namely, the scales have a hierarchy
\begin{eqnarray}
({\rm string})&\underset{M_s}{\longrightarrow}& {\rm
(d=4+\delta,\ N=1\ supergravity)}
\nonumber\\
&\underset{\MG}{\longrightarrow}& {\rm (d=4,\ N=1\ supergravity\ MSSM)} 
\end{eqnarray}
Thus, toward a 4D observer at low energy there arise towers
of KK states above the scale $\MG$. 
Including these states in the running of gauge couplings
between $\MG$ and $M_s$, we obtain\cite{dienes}
\begin{eqnarray}
\alpha_a^{-1}(\Lambda)=\alpha_a^{-1}(\mu_0)
-\frac{b_a-\tilde b_a}{2\pi}\ln\frac{\Lambda}{\mu_0}
-\frac{\tilde b_a X_\delta}{2\pi\delta}\left[\left(
\frac{\Lambda}{\mu_0}\right)^\delta-1\right]
\end{eqnarray}  
where $b_a$ is the beta function coefficient of the
group $G_a$ contributed by all the MSSM fields,
$\tilde b_a$ is the one contributed by the bulk fields.
Keeping the power law divergent term only
in asymptotically free models, we obtain
\begin{equation}
\frac{\MG}{M_s}\longrightarrow
\frac{\mu_0}{\Lambda}\simeq \left(
\frac{-\tilde b X_\delta \alpha_a(\Lambda)}{2\pi\delta}
\right)^{1/\delta}
\end{equation}
The KK sum is from the lowest one $1/R$ to the highest
one $M_s$. Thus, the length scale describing the internal
space is $R\sim 1/\MG$ and the string scale is $1/M_s$, giving
a ratio of the compactification volumes as 
\begin{equation}
\frac{\MG}{M_s}\sim \left(\frac{ 
L_s}{R}\right)
\end{equation} 
which is another way of saying that the small number
(\ref{small})
needs a large volume in the extra $\delta$-dimension.
For the Friedmann-Witten case (\ref{g2}), we do not obtain
this relation exactly even though a large volume effect 
must be there.

The gaugino masses evolve as
\begin{equation}\label{gaugino}
M_a(\MG)=\left(\frac{g_a(\MG)}{g_a(M_s)} \right)^2 M_a
(M_s)\underset{\tilde b<0}{\longrightarrow} 
\frac{M_a(\MG)}{M_a(M_s)}
\simeq\frac{C_2(G_a) X_\delta
\alpha_{GUT}}{\pi\delta}
\left(\frac{M_s}{\MG}\right)^\delta,
\end{equation}
and the soft scalar masses evolve as\footnote{It is
required that the bulk matter Yukawa couplings do not
dominate over the bulk gauge running. Without bulk 
matter\cite{kubo}, this condition is satisfied.
In the next section, we show that even bulk matter with 
top quark coupling is not harmful.}
\begin{equation}\label{soft}
(m^2)^i_j(\MG)= \frac{C_2(R_i)}{C_2(G_a)}
\left[1-\left(\frac{g_a(M_s)}{g_a(\MG)}
\right)^4 \right] \delta^i_j |M_a(\MG)|^2
+(m^2)^i_j(M_s)
\end{equation}
where $C_2$ is the quadratic Casimir operator, e.g.
$C_2(SU_5)=5, C_2({\bf 5})=\frac{12}{5}$ and $C_2({\bf 10})
=\frac{18}{5}$.
From the gaugino mass evolution (\ref{gaugino}), we note that
due to the small number (\ref{small})  
$M_a(\MG)/M_a(M_s)$ is large if the group $G_a$ is asymptotically
free. Then, if $(m^2)^i_j$ is small or comparable to the
gaugino mass squared at $M_s$, the soft mass term at the
scale $\MG$ is dominated by the diagonal element due to
Eq. (\ref{soft}). This relative enhancement of the soft
mass from the KK mode contribution in the bulk is
the KT scenario of suppressing the flavor changing neutral
current. For this scenario to work, {\it one needs an
asymptotically free gauge interaction in the bulk
and a large internal space volume.}
In this case, it was also pointed out\cite{kubo} that the soft $A$ 
and $B$ terms tend to the flavor conserving points. 
We confirm that these conclusions are true.

As a numerical guide, we present the evolution of
the gaugino mass in 5D in Fig. 1
with all Yukawa couplings set at zero except that
of the top quark. We chose a figure with 1.5 TeV gluino
mass at $M_Z$. Here, we assumed the MSSM spectrum 
between $M_Z$ and $\MG$. 
We use the gluino mass at $M_Z$ as an input.
For the observed values of the
strong, weak and electromagnetic coupling constants, 
we run the parameters up to $\MG$. 
The wino and zino(or photino) masses
are obtained by running the unified gaugino mass(the
gluino mass) from
$\MG$ down to $M_Z$. Between $\MG$ and $M_s$, a SUSY
$SU(5)$ is assumed with three families and two Higgs
quintets but without a ${\bf 24}_H$.[A geometrical
breaking of $SU(5)$ does not need an adjoint Higgs
field.]  
Indeed, $M_5(\MG)/M_5(M_P)$
is a big number and softening of the flavor changing
neutral current is achieved.   
If we had worked in 6D as in Ref.\cite{kubo}, it 
would not be a consistent calculation since it has a 6D gauge
anomaly. 

\vskip 0.3cm
\begin{figure}
\begin{center}
\includegraphics[angle=-90,width=0.8\textwidth,clip=]{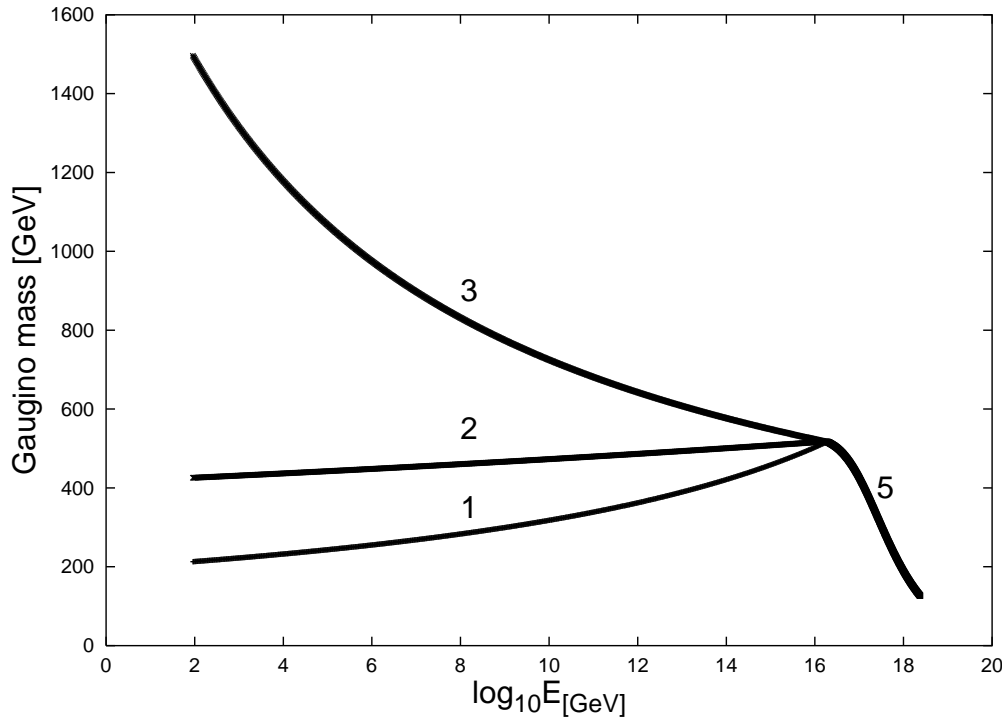}
\vskip 0.6cm
\caption{The evolution of gaugino masses with a 5D $SU(5)$ GUT.
$SU(N)$ and $U(1)$ gauginos are marked by $N$ and 1,
respectively.}
 {\label{fig:gaugino mass}}
\end{center}
\end{figure}
\vskip 0.3cm

\section{Bulk matter contribution}

For the KT scenario to work, the Yukawa couplings
should not behave in the same way as the soft mass
behavior discussed in Sec. II. 
Namely, the Yukawa
couplings should not be diagonalized so that a reasonable
quark mixing matrix is obtained. 
If there is no bulk matter\cite{kubo}, this differentiation
is easily achieved. The main reason for this differentiation is 
that the soft masses are renormalized additively
but the main contribution to the Yukawa coupling
renormalization is multiplicative. It is
ironic to observe that {\it 
the very nature of the additive renormalization
of scalar masses needed supersymmetry for the gauge hierarchy 
solution, but the SUSY flavor problem created by 
supersymmetrization employs this additive 
renormalization property toward a
solution of the SUSY flavor problem.}

On the other hand, if we introduce bulk matter, the discussion
is more involved, which we show below. 
For the bulk matter, we have the following 
renormalization group equation for the Yukawa coupling $Y^{ijk}$
\cite{kubo2}
\begin{eqnarray}
\Lambda\frac{d Y^{ijk}}{d\Lambda}&=&
\frac{1}{32\pi^2}\left( Y^{ijl}Y^{kmn}
+(j\leftrightarrow k)+(i\leftrightarrow k)\right)
Y_{lmn} \nonumber \\
&&+\frac{1}{32\pi^2}\sum_{\{l,m,n\}}\biggl[ X_{\delta} \left(
\frac{\Lambda}{\MG} \right)^\delta-1\biggr]
\left( Y^{ijl}Y^{kmn}+(j\leftrightarrow k)
+(i\leftrightarrow k)\right)Y_{lmn} \nonumber \\
&&-\frac{2}{16\pi^2}\left( C_2(R_i)+C_2(R_j)
+C_2(R_k)\right)G^2_\delta Y^{ijk}\label{Yrunning} 
\end{eqnarray}
where the dummy index sum is for bulk matter $\{l,m,n\}$, and
\begin{equation}
G_\delta^2=g^2 X_\delta\left(\frac{\Lambda}{\MG}\right)^\delta
\end{equation}
which becomes constant due to the asymptotically free
gauge coupling. The first term of Eq. (\ref{Yrunning}) 
comes from the zero modes(brane and bulk) 
of the fields in the loop, the second term comes from 
the diagrams which contain KK modes in the loop. 
The third term of Eq. (\ref{Yrunning}) comes 
from the diagrams which contain 
gauge fields in the loop(the zero mode and KK modes). 
Since the first term is logarithmic it can be ignored 
compared to the other two terms. The second and third terms
are the power-running and have different signs. According 
to the relative magnitudes, the Yukawa coupling can increase or 
decrease as $\Lambda$ increases. 
Thus, let us compare the magnitudes of these two terms as 
the energy scale $\Lambda$ increases from \Mg\ to $M_s$.

These two terms have the following relative magnitudes,
\begin{eqnarray}
&&28[ X_{\delta} \left(\frac{\Lambda}{\MG} 
\right)^\delta-1\biggr]Y^2 \label{object1}\\
&&-12C_2(r)G^2_\delta.  \label{object2}
\end{eqnarray}
The coefficient 28 in (\ref{object1}) is obtained for $E_6$
Yukawa coupling ${\bf 27\cdot 27\cdot 27}_H$ with the
third family coupling strength. 
We included the color factor and contributions of
${\bf 10}\subset {\bf 27}$ of the loop particles.
For a simple numerical comparison, we assumed
$C_2(R_i)=C_2(r)$ in Eq. (\ref{object2}). We set $\delta=2$, i.e.
$D=6$ and $X_2=\pi$.

For Eq. (\ref{object1}), we consider first $\Lambda\simeq\MG$, and
then consider for a general $\Lambda$. 
For example, consider one {\bf 27} of $E_6$ in the bulk. Then,
for Eqs. (\ref{object1}) and (\ref{object2})
we obtain at $\Lambda\simeq\MG$, 
\begin{equation} 
28[\pi-1]Y^2\simeq 29.4 
\end{equation}
and 
\begin{equation}
-12 C_2({\bf 27}) G_2^2=-163,
\label{c2}
\end{equation}
respectively, using $Y\simeq 0.7$. In Eq. (\ref{c2}), we used
$G_\delta$ evaluated at $\Lambda=\MG$. 
Thus, near the scale $\MG$ Eq. (\ref{object2}) is the dominant
term and the Yukawa coupling decreases as the scale increases.
For $Y^2(\Lambda)$ evaluation, we need
\begin{equation}
\left(\frac{g(M_{GUT})}{g(\Lambda)} \right)^2
=1+\frac{C_2(G)X_\delta g^2(M_{GUT})}{4\pi^2\delta} 
\left\{ \left( \frac{\Lambda}{M_{GUT}}\right)^\delta-1 \right\}.
\label{ratioofg}
\end{equation}
Then, $Y^2(\Lambda)$ is
\begin{eqnarray}
Y^2(\Lambda)&=&\left(\frac{g(\Lambda)}{g(\MG)}
\right)^{2\eta^{ijk}_Y}Y^2(\MG)\nonumber\\
&=&\biggl[1+\frac{C_2(G)X_\delta g^2(M_{GUT})}{4\pi^2\delta} 
\left\{ \left( \frac{M_P}{\MG}\right)^\delta-1 
\right\} \biggr]^{-\eta^{ijk}_Y}Y^2(\MG)\\
&\simeq&\biggl[1+\frac{6}{25}\left\{\left( 
\frac{\Lambda}{\MG}\right)^2-1\right\}\biggr]^{-13/6}
\left(0.7\right)^2\nonumber
\end{eqnarray} 
where $\eta_Y^{ijk}=(C_2(R_i)+C_2(R_j)+C_2(R_k))/C_2(G_a)$.
For the last equality, we used $C_2(E_6)=12$ and $Y(\MG)=0.7$.
Thus, Eq. (\ref{object1}) becomes
\begin{equation}
28 \biggl[ X_{\delta} \left(\frac{\Lambda}{\MG} 
\right)^2-1\biggr]\biggl[1+\frac{6}{25}
\left\{\left( \frac{\Lambda}{\MG}\right)^2-1
\right\}\biggr]^{-13/6}\left(0.7\right)^2
\label{lambda}
\end{equation}
which can be compared to the magnitude given in (\ref{object2})
which is saturated to
\begin{equation}\label{sat}
-96\pi^2\frac{C_2({\rm{\bf 27}})}{C_2({\bf 78})}=-684.
\end{equation}
The maximum value of (\ref{lambda}) is about 54 
$\simeq 2\MG$. Note the value of Eq. (\ref{object2}) is 
--384 at $\Lambda \simeq 2\MG$. Therefore, Eq. (\ref{object1}) 
can never exceed the magnitude
of (\ref{object2}). Therefore,
Eq. (\ref{object2}) continues to dominate Eq. (\ref{object1}).
Hence, we can approximate the Yukawa coupling running
given by the third term of Eq. (\ref{Yrunning}), leading to
\begin{equation}
Y^{ijk}(\MG)=\left(\frac{g(\MG)}{g(M_s)}
\right)^{\eta^{ijk}_Y}Y^{ijk}(M_s)
\end{equation}
This is a multiplicative result and the needed inter-family
mixings are not suppressed.

For the other soft terms($m^2,A$ and $B$ terms)
we have checked the
contributions of $Y^2$ to the evolution equations are
also negligible, and bulk matter can be allowed
toward a successful KT mechanism.

\section{The anomaly}

In the previous section,
the softening of the SUSY flavor problem has been
obtained by a large number of $M_a(\MG)/M_a(M_s)$
which depends on $[g(\MG)/g(M_s)]^2$ which in turn
depends on $\sim (M_s/\MG)^\delta$. Given the small
number of order $10-100$ for $M_s/\MG$, a larger
$\delta$ can remove the unwanted flavor violating
pieces more effectively. We argure, in accord with Kubo and 
Terao, that $\delta=1$ is not sufficient.\footnote{If $\delta=1$,
some flavor changing problems can be evaded 
but the SUSY CP problem is
difficult to understand\cite{masiero} 
with the extra dimensional scenario
alone. However, if $1/R\ll \MG$ is assumed, $\delta=1$
can be admissible.} 
Thus, we consider $\delta=2$, i.e. 6D SUSY field
theories. 

Then, we search for models satisfying two
conditions: (i) no gauge anomaly, and (ii) asymptotic
freedom in the bulk. 

One should consider also the gravitational anomaly\cite{agw},
but it is easy to remove it by adding gauge singlet fermions.
Thus, we will not use the vanishing 
gravitational anomaly as an absolute condition.

Note that there exist square anomalies in 
6D\cite{anomaly6d}. We are interested in the $A,D,E$ series. 
The asymptotic freedom condition is calculated from the
fields in the bulk. The gauge multiplet splits into
an $N=1$ gauge multiplet plus a chiral multiplet in 4D,
and a hyper multiplet splits into two chiral multiplets
with opposite quantum numbers in 4D. Thus, we require 
\begin{equation}\label{condasym}
-2C_2(G)+\sum_i 2\ell(R_i)<0
\end{equation}
where $\ell(R_i)$ is the index of the representation $R_i$,
and the sum is for the bulk hyper multiplet representations.

\vskip 0.2cm
\subsection{\bf $SU(N)$ and $SO(2n)$}

The groups $SU(N)$ and
$SO(2n)$ have the following anomalies for the same chirality
fermions,
\begin{eqnarray}
SU(N)&:&2N\ ({\rm for \ adjoint}),\nonumber\\
&\ & \frac{(N-4)![N(N+1)-
6j(N-j)]}{(j-1)!(N-1-j)!}\ ({\rm for\ }j\ {\rm antisymmetric\ 
indices\ }[j]) \label{suN}\\
&\ &1\ ({\rm for\ fundamental})\nonumber
\\
SO(2n)&:&4(n-4)\ ({\rm for\ adjoint}),\nonumber\\
 &\ & -2^{n-4}\ ({\rm for\ 
spinor}),\label{so2n}\\
 &\ &2\ ({\rm for\ fundamental})\nonumber
\end{eqnarray}
Note, however, that the vector multiplet and hyper multiplets
have the opposite chiralities to be consistent with 
supersymmetry. 
Most models satisfying the anomaly free condition
do not satisfy the asymtotic freedom condition. Note that
among the models presented in Ref.\cite{yanagida} the $SU(5)$
(ten {\bf 5} in the bulk) is not allowed but $SO(10)$
(three {\bf 10}'s and one $\overline{\bf 16}$) 
is allowed.

We have not found any 6D $SU(N)$ model satisfying the
two conditions. In $SO(10)$, there are two models:
one model with two {\bf 10}'s and the other with
three {\bf 10}'s and one {\bf 16}(or $\overline{\bf 16}$).
For $SO(2n)$ with $n\ge 6$, there always exists one
model: $2(n-4)$ fundamental representations.

\vskip 0.2cm
\subsection{\bf Exceptional groups}

But more interesting groups are the exceptional groups.
It is known that the exceptional groups are anomaly free
in 6D\cite{anomaly6d}.
In this sense, the exceptional groups in the $E$-series
can be claimed to be the grand unification groups in 6D, 
as the orthogonal groups in the $D$-series are
considered as the grand unfication groups in 4D. 
In fact an anomaly-free
$E_7$ 6D model was obtained by a $Z_2$ orbifold 
compactification of the heterotic string\cite{6d}.
Its $E_7$ spectrum is one ${\bf 133}$(gauge multiplet)
plus ten {\bf 56}'s(hyper multiplets). But these do not
satisfy the asymptotic freedom condition in the bulk.

Here, we consider any 6D $E_6,E_7,E_8$ field 
theoretic models
in the bulk, hoping that they can be obtained from some 
compactification of string models. For these, we obtain 
constraints for the number of fundamental representations
in the bulk matter
\begin{eqnarray}
E_6&:& n_{27}\le 3 
\nonumber\\
E_7&:& n_{56}\le 2 
\\
E_8&:& {\rm no\ hyper\ multiplet\ in\ the\ bulk}
\nonumber
\end{eqnarray}
where we used $C_2(E_i)= (12, 18, 30)$ for $(i=6, 7, 8)$, 
respectively, 
and $\ell({\bf 27}_{E_6})= 3$ and $\ell({\bf 56}_{E_7})=6$.
Thus, in principle it is possible to put all chiral matter
representations in the bulk. 

It will be of utmost importance to search for 6D SUSY
models with the above property through the string 
compactifications. They can be considered as the string
solutions of the SUSY flavor problem.

\section{Conclusion}

We have considered 6D groups with the
asymptotic freedom in the bulk, toward softening of the
SUSY flavor problem. We extended the KT mechanism to
include the possibility of hyper multiplets in the bulk.
The two conditions on the anomaly freedom and
asymptotic freedom exclude most 6D GUT models,
except two $SO(10)$ models, $SO(2n)$ 
models(one each for each $n\ge 6$), 
and several exceptional group models.
The suppression of the flavor changing neutral current
is obtained because of the existence of a small
number (\ref{small}) due to a large internal 6D
volume compared to the Planckian volume(or the string
scale volume).

\acknowledgments

We thank Hyun Min Lee and Hans Peter Nilles for useful
discussions.
One of us(JEK) thanks Humboldt Foundation for the award.
This work is supported in part by the BK21 program of Ministry 
of Education, KOSEF Sundo Grant, and 
Korea Research Foundation Grant No. KRF-PBRG-2002-070-C00022.

\end{document}